\documentclass[12pt,a4paper]{article}

\newcommand{\la}{\lambda}
\newcommand{\pa}{\partial}
\newcommand{\te}{\tau_e}
\newcommand{\al}{\alpha}
\newcommand{\rz}{\rho_0}
\newcommand{\oa}{\omega_+}
\newcommand{\ob}{\omega_-}

\begin{document}

\begin{flushright}
{ }
\end{flushright}
\vspace{1.8cm}

\begin{center}
 \textbf{\Large Algebraic Curves for Long Folded and \\
Circular Winding Strings in $AdS_5\times S^5$}
\end{center}
\vspace{1.6cm}
\begin{center}
 Shijong Ryang
\end{center}

\begin{center}
\textit{Department of Physics \\ Kyoto Prefectural University of Medicine
\\ Taishogun, Kyoto 603-8334 Japan}
\par
\texttt{ryang@koto.kpu-m.ac.jp}
\end{center}
\vspace{2.8cm}
\begin{abstract}
For the homogeneous configuration given by the long string limit of
the folded string with a spin in $AdS_3$ and a spin and a winding number
in $S^1$, we solve the auxiliary linear problem in the finite-gap
method and construct the Lax operator to obtain an algebraic curve.
We show that the long spiky string in $AdS_3$ has the same algebraic
curve as the null cusp Wilson loop. The algebraic curve for the circular
winding string in $AdS_3\times S^1$ is evaluated. The Virasoro constraint
is discussed to characterize the branch points of each curve.
\end{abstract}
\vspace{3cm}
\begin{flushleft}
December, 2012 
\end{flushleft}

\newpage
\section{Introduction}

The AdS/CFT correspondence \cite{MGW} has more and more revealed the deep
relations between the $\mathcal{N}=4$ super Yang-Mills theory
and the string theory in $AdS_5 \times S^5$ \cite{NB}. Various types of
classical string solutions play an important role for the computation of
the planar contribution to the conformal dimensions of non-BPS operators
for any value of the coupling constant and the construction of the 
correlation functions for string states \cite{JSW,KM,SR,JW}.

The finite-gap method for spectrum of classical string theory has been
presented \cite{KMM,KZ,BKS,SN} where each solution of string in 
$AdS_5 \times S^5$ is characterized by a Riemann surface known as the
spectral curve. This approach has been developed \cite{DV,SS,ND,BV,DL},
where various string solutions and their dynamics are formulated 
as the finite-gap solutions.
The spiky string solution in $AdS_3$ \cite{MK,JJ} has been investigated
by the finite-gap method in \cite{DL} and the spectral curve 
 defined \cite{KMM,KZ} by using the monodromy operator has been
studied where the eigenvalue of the monodromy operator
specifies the pseudomomentum whose analytic properties as a function 
of the spectral parameter are analyzed. 
The algebraic curve has been defined by using the Lax operator
which is described  as the derivative with respect to the
spectral parameter for the logarithm of the 
monodromy operator \cite{DV,NKS}.

The finite-gap solution in \cite{KMM} has been produced by investigating
the analytic properties of monodromy around a noncontractible loop going
around the closed string cylinder. In ref. \cite{JG} the algebraic curves
based on the Lax operator have been constructed for the Wilson loop 
minimal surfaces corresponding to the null cusp \cite{NKR,AM} and 
$q\bar{q}$ potential \cite{CHR}, where all loops are contractible.
For the classical string solution \cite{KZA}
dominating the correlation function of
the circular Wilson loop and a local operator, an algebraic curve has
been also presented. Further inversely the original Wilson loop minimal
surface and the correlation function solution have been reconstructed from
the algebraic curves and some minimal structural assumptions.

Associated with the finite-gap method and the Polhmeyer reduction 
method \cite{KP} there have been various constructions of string solutions
by using theta functions \cite{HOS,KSS,IKZ} 
and evaluations of the three-point
correlators for three heavy string states \cite{JW,KK}.

The null cusp Wilson loop solution is related \cite{KRT} with the large
spin limit of the GKP folded string with spin $S$ in $AdS_3$ \cite{GKP}
through a discrete SO(2,4) transformation and an analytic continuation.
In the special large spin scaling limit, namely the long string limit
\cite{BGK,FTT} the folded closed string with two spins $S$ in $AdS_3$
and $J$ in $S^1$ turns out to be a homogeneous configuration.
The one- and two-loop corrections to the energy of this string 
configuration have been computed \cite{FTT,RTT,RT} and compared with
the cusp anomalous dimension derived from the strong coupling
behavior of the asymptotic Bethe 
ansatz \cite{BES,CK,BKK} and the O(6) sigma model analysis \cite{LAM}.
Further the folded string with spin $S$ in $AdS_3$ and spin $J$ and
winding number $m$ in $S^1$ becomes a generalized homogeneous 
configuration in the large spin scaling limit.
The two-loop correction to the energy of this string configuration
has been studied \cite{GRR} to be in agreement with the corresponding 
terms in the generalized scaling function as obtained from the asymptotic
Bethe ansatz \cite{NG} and also partially from the quantum O(6) sigma 
model and the Bethe ansatz data \cite{BBB}. 

Using the procedure of ref. \cite{JG} we will construct the algebraic 
curve for the homogeneous string solution with two spins $S$ and $J$
in $AdS_3 \times S^1$ \cite{FTT} as well as for the generalized 
homogeneous string solution  with an additional winding
number $m$ in $S^1$ \cite{GRR}. It is desirable to study how the
corresponding algebraic curves are related with the curve for
the null cusp Wilson loop solution. For the long spiky string solution
in $AdS_3$ \cite{DL}, the algebraic curve based on the Lax operator
will be computed. We will analyze the circular winding string solution 
with spin $S$ and winding number $n$ in $AdS_3$ and with spin $J$ and 
winding number $m$ in $S^1$ \cite{ART} to derive the algebraic curve.

\section{Algebraic curve for the long folded string with two spins $S$ and
$J$}

Based on the prescription in ref. \cite{JG} we express the action for the
$AdS_3$ sigma model in terms of group elements
\begin{equation}
S = \frac{\sqrt{\la}}{4\pi}\int \mathrm{tr} j\bar{j} d^2w,
\end{equation}
where $w$ and $\bar{w}$ are the complex worldsheet coordinates and the
currents $j = g^{-1}\pa g$ and $\bar{j} = g^{-1}\bar{\pa} g$ 
are specified by the group element
\begin{equation}
g = \left( \begin{array}{cc} \frac{x_0 + x_1}{z} & \frac{1}{z} \\
-\frac{-x_0^2 + x_1^2 + z^2}{z} & \frac{x_0 - x_1}{z} \end{array}  \right)
\label{ge}\end{equation}
for the Minkowskian $AdS_3$ in the Poincare coordinates $(z_0,x_1,x_2)$.
In the finite-gap method the equations of motion  
 are described by the zero-curvature condition for a 
family of flat currents represented by the spectral parameter $x$ as
\begin{equation}
J = \frac{j}{1 - x}, \hspace{1cm} \bar{J} = \frac{\bar{j}}{1 + x}.
\end{equation}

The zero-curvature condition enforced for arbitrary complex $x$ is 
equivalent to the consistency condition for the auxiliary linear problem
\begin{eqnarray}
\pa \Psi &+& J\Psi = 0, 
\label{lp} \\
\bar{\pa} \Psi &+& \bar{J}\Psi = 0, 
\label{bl}\end{eqnarray}
where $\Psi(w,\bar{w};x)$ is a two-component vector. A $2\times 2$ matrix
$\hat{\Psi}(w,\bar{w};x)$ is constructed by arranging the two independent
solutions $\Psi_1^a, \Psi_2^a$ into two column vectors as 
\begin{equation}
\hat{\Psi}(w,\bar{w};x) = (\Psi_1, \Psi_2),
\end{equation}
which obeys the matrix differential equations
\begin{equation}
\pa \hat{\Psi} + J\hat{\Psi} = 0, \hspace{1cm} \bar{\pa} \hat{\Psi} + 
\bar{J}\hat{\Psi} = 0.
\end{equation}
The solutions of these equations lead to the current
\begin{equation}
j = -\pa \hat{\Psi}\cdot{\hat{\Psi}}^{-1}|_{x=0},
\end{equation}
which reproduces the original string solution \cite{DV}
\begin{equation}
g = \sqrt{\det \hat{\Psi}}\cdot {\hat{\Psi}}^{-1}|_{x=0}.
\label{gx}\end{equation}

From the monodromy operator expressed in terms of the flat currents as
\begin{equation}
\Omega(w_0,\bar{w_0};x) = Pe^{\int_C Jdw + \bar{J}d\bar{w}},
\end{equation}
where $C$ is a loop on the worldsheet based on some point 
$(w_0,\bar{w_0})$, a Lax operator is defined by
a $2\times 2$ matrix \cite{DV}
\begin{equation}
L(w,\bar{w};x) = -i \frac{\pa}{\pa x}\log \Omega(w,\bar{w};x),
\end{equation}
whose entries are rational ( or polynomial ) functions of $x$.
The Lax operator satisfies
\begin{equation}
\pa L + [J, L] = 0, \hspace{1cm} \bar{\pa} L + [\bar{J}, L] = 0
\label{el}\end{equation}
and defines an algebraic curve as the characteristic equation
\begin{equation}
\det(y\cdot 1 - L(w,\bar{w};x)) =0.
\label{cu}\end{equation}
The solution of (\ref{el}) is presented as \cite{JG}
\begin{equation}
L(w,\bar{w};x) = \hat{\Psi}(w,\bar{w};x)\cdot A(x)\cdot
\hat{\Psi}(w,\bar{w};x)^{-1},
\label{la}\end{equation}
where $A(x)$ is an arbitrary $x$-dependent matrix.

We consider the folded closed string with spin $S$ in $AdS_3$ \cite{GKP}
and spin $J$ in $S^1$ \cite{FT}. In the long string limit
\begin{equation}
S \gg J \gg 1, \hspace{1cm} \hat{\nu} \equiv \frac{\nu}{\mu} 
 \approx \frac{\pi J}{\sqrt{\la}\ln S} = \mathrm{fixed},
\end{equation}
the $(S,J)$ folded string solution turns out to be the following 
homogeneous configuration in $AdS_3\times S^1$ \cite{FTT}
\begin{eqnarray}
ds^2 &=& - \cosh^2\rho dt^2 + d\rho^2 + \sinh^2\rho d\phi^2 + d\varphi^2,
\label{me} \\
t &=& \kappa \tau, \;\; \rho = \mu \sigma, \;\; \phi = \kappa \tau, \;\;
\varphi = \nu \tau, \;\; \kappa, \mu, \nu \gg 1,
\label{ho}\end{eqnarray}
where the parameters $\mu$ and $\nu$ are related with spins $S$ and $J$ as
\begin{equation}
\mu \approx \frac{1}{\pi}\ln S, \hspace{1cm} \nu = \frac{J}{\sqrt{\la}}.
 \end{equation}
The Virasoro constraint gives 
\begin{equation}
\kappa^2 = \mu^2 + \nu^2.
\label{vi}\end{equation}
The corresponding embedding coordinates in $AdS_3$ are described by
\begin{equation}
Y_{-1} + iY_0 = \cosh\mu\sigma e^{i\kappa\tau}, \hspace{1cm} 
Y_1 + iY_2 = \sinh\mu\sigma e^{i\kappa\tau}.
\end{equation}

We rotate the worldsheet time $\tau$ to the Euclidean one
\begin{equation}
\tau = -i \te
\label{an}\end{equation}
to have 
\begin{eqnarray}
Y_{-1} &=& \cosh\mu\sigma \cosh\kappa\te, \hspace{1cm} 
Y_0 =-i\cosh\mu\sigma \sinh\kappa\te, \nonumber \\
Y_1 &=& \sinh\mu\sigma \cosh\kappa\te, \hspace{1cm} 
Y_2 =-i\sinh\mu\sigma \sinh\kappa\te.
\end{eqnarray}
Performing a discrete SO(2,4) transformation that interchanges $Y_0$ and
$Y_2$ as $Y_0 = -iY_2', \; Y_2 = -iY_0'$ we express the string solution in
terms of the Minkowskian Poincare coordinates 
\begin{eqnarray}
\frac{1}{z} &=& Y_{-1} - Y_2' = e^{-\kappa\te}\cosh\mu\sigma, \nonumber \\
x_0 &=& zY_0' = e^{\kappa\te}\sinh\kappa\te \tanh\mu\sigma, \;\;
x_1 = zY_1 = e^{\kappa\te}\cosh\kappa\te \tanh\mu\sigma.
\end{eqnarray}
It implies that the group element in (\ref{ge}) is given by
\begin{equation}
g = \left( \begin{array}{cc} e^{\kappa\te}\sinh\mu\sigma &
e^{-\kappa\te}\cosh\mu\sigma  \\ -e^{\kappa\te}\cosh\mu\sigma 
& -e^{-\kappa\te}\sinh\mu\sigma \end{array} \right).
\end{equation}
For the Euclidean worldsheet we represent the complex coordinates as
\begin{equation}
w = \sigma + i\te, \hspace{1cm}  \bar{w} = \sigma - i\te
\label{st}\end{equation}
to have the currents
\begin{eqnarray}
j &=& \left( \begin{array}{cc} -i\frac{\kappa}{2} & 
\frac{\mu}{2}e^{i\kappa(w-\bar{w})} \\ 
\frac{\mu}{2}e^{-i\kappa(w-\bar{w})} & i\frac{\kappa}{2}
\end{array} \right), 
\label{jh} \\
\bar{j} &=& \left( \begin{array}{cc} i\frac{\kappa}{2} & 
\frac{\mu}{2}e^{i\kappa(w-\bar{w})} \\ 
\frac{\mu}{2}e^{-i\kappa(w-\bar{w})} & -i\frac{\kappa}{2}
\end{array} \right).
\label{bj}\end{eqnarray}

Expressing the two-component vector $\Psi$ as $\Psi= (\begin{array}{c}
\al \\ \beta \end{array} )$ for the linear differential equation 
(\ref{lp}) we get the following two equations
\begin{eqnarray}
\left( \pa + \frac{1}{1-x}j_{22} \right)\left( \pa + \frac{1}{1-x}j_{11}
\right)\al &-& \frac{1}{(1-x)^2}j_{12}j_{21}\al = 
\frac{\pa j_{12}}{j_{12}} \left( \pa + \frac{1}{1-x}j_{11}
\right)\al,
\label{al} \\
\left( \pa + \frac{1}{1-x}j_{22} \right)\beta &=& 
- \frac{1}{1-x}j_{21}\al.
\label{be}\end{eqnarray}
The substitution of each entry of the current $j$ into (\ref{al})
leads to two independent solutions
\begin{equation}
\al_{\pm} = e^{\frac{\kappa}{2}\left( i \pm \frac{1}{1-x}
\sqrt{\left( \frac{\mu}{\kappa} \right)^2 - x^2} \right)w }.
\label{pm}\end{equation}
We choose $\al_{-}$ as associated with the upper component of $\Psi_1$
and obtain the corresponding lower component $\beta_-$ by combining
(\ref{be}) and (\ref{pm})
\begin{equation}
\beta_- = \frac{\kappa}{\mu}\left(\sqrt{\left( 
\frac{\mu}{\kappa} \right)^2 - x^2} + ix \right) 
 e^{\frac{\kappa}{2}\left( -i - \frac{1}{1-x}
\sqrt{\left( \frac{\mu}{\kappa} \right)^2 - x^2} \right)w }.
\end{equation}

The other linear differential equation (\ref{bl}) is also 
expressed as
\begin{eqnarray}
\left( \bar{\pa} + \frac{1}{1+x}\bar{j_{22}} \right)\left( \bar{\pa} + 
\frac{1}{1+x}\bar{j_{11}}\right)\al &-& \frac{1}{(1+x)^2}\bar{j_{12}}
\bar{j_{21}}\al = \frac{\bar{\pa} \bar{j_{12}}} {\bar{j_{12}}} 
\left( \bar{\pa} + \frac{1}{1+x}\bar{j_{11}}\right)\al,
\nonumber \\
\left( \bar{\pa} + \frac{1}{1+x}\bar{j_{22}} \right)\beta &=& 
- \frac{1}{1+x}\bar{j_{21}}\al,
\label{bb}\end{eqnarray}
which characterizes the $\bar{w}$ dependence.
According to the minus sign of the square root we extract 
the $\bar{w}$-dependent factor
\begin{eqnarray}
\tilde{\al_-} &=& e^{\frac{\kappa}{2}\left( -i - \frac{1}{1+x}
\sqrt{\left( \frac{\mu}{\kappa} \right)^2 - x^2} \right)\bar{w} },
\nonumber \\
\tilde{\beta_-} &=& e^{\frac{\kappa}{2}\left( i - \frac{1}{1+x}
\sqrt{\left( \frac{\mu}{\kappa} \right)^2 - x^2} \right)\bar{w} }.
\end{eqnarray}
Combining together we obtain one independent solution
$\Psi_1$ for the linear problem
\begin{equation}
\Psi_1 = e^{-\frac{\kappa}{2}\sqrt{\left( \frac{\mu}{\kappa} \right)^2
 - x^2} \left( \frac{1}{1-x}w +  \frac{1}{1+x}\bar{w} \right)}
\left( \begin{array}{c} e^{i\frac{\kappa}{2}(w - \bar{w})} \\
\frac{\kappa}{\mu}\left(\sqrt{\left( \frac{\mu}{\kappa} \right)^2 - x^2} +
 ix \right) e^{-i\frac{\kappa}{2}(w - \bar{w})} \end{array}\right).  
\label{so}\end{equation}

In the same way the other independent solution $\Psi_2$ corresponding to
the plus sign choice in (\ref{pm}) is also constructed by
\begin{equation}
\Psi_2 = e^{\frac{\kappa}{2}\sqrt{\left( \frac{\mu}{\kappa} \right)^2
 - x^2} \left( \frac{1}{1-x}w +  \frac{1}{1+x}\bar{w} \right)}
\left( \begin{array}{c} e^{i\frac{\kappa}{2}(w - \bar{w})} \\
\frac{\kappa}{\mu}\left(-\sqrt{\left( \frac{\mu}{\kappa} \right)^2 - x^2}+
 ix \right) e^{-i\frac{\kappa}{2}(w - \bar{w})} \end{array}\right).  
\label{ts}\end{equation}
Thus the two independent solutions differ by choosing a different 
branch of the square root, where we treat the linear solution as a
single function on the two branches of the algebraic curve.
The solutions (\ref{so}) and (\ref{ts}) show the isolated essential 
singularities at $x = \pm 1$, which is the structure of the
Baker-Akhiezer function \cite{JG}.

From the expression (\ref{la}) with (\ref{so}) and (\ref{ts}) we obtain
the Lax matrix which is generally not a polynomial in $x$.
We choose $A(x)$ to be
\begin{equation}
A(x) = \sqrt{\left( \frac{\mu}{\kappa} \right)^2 - x^2}\left( 
\begin{array} {cc} 1 & 0 \\ 0 & -1 \end{array} \right),
\end{equation}
which gives the polynomial Lax matrix
\begin{equation}
L(w,\bar{w};x) = \left( \begin{array} {cc} -ix & \frac{\mu}{\kappa}
e^{i\kappa(w-\bar{w})} \\ \frac{\mu}{\kappa}e^{-i\kappa(w-\bar{w})} &
ix \end{array} \right).
\end{equation}
The substitution of this expression into (\ref{cu}) yields the genus-0
algebraic curve
\begin{equation}
y^2 = \left( \frac{\mu}{\kappa} \right)^2 - x^2.
\label{ch}\end{equation}
Owing to the Virasoro constraint (\ref{vi}) the branch points
of the curve $x^{\pm} = \pm \mu/\kappa$ stay 
between the essential singular points
$\pm 1$ as $-1 < x^- < x^+ < 1$.
The algebraic curve is also expressed in terms of the fixed 
scaling parameter $\hat{\nu}$ 
\begin{equation}
y^2 = \frac{1}{1 + {\hat{\nu}}^2 } - x^2.
\end{equation} 
When the spin $J$ in $S^1$ is turned off, that is, $\kappa = \mu$ 
the algebraic curve becomes $y^2 = 1 - x^2$, which is the same as
the expression \cite{JG} for the null cusp Wilson loop solution
\cite{NKR,AM}.

\section{Algebraic curve for the long spiky string in $AdS_3$}

We turn to the spiky string solution \cite{MK} in $AdS_3$ for the 
string sigma model in the conformal gauge \cite{JJ,DL}.
In the large spin limit the spiky string solution in the metric (\ref{me})
is described by \cite{DL}
\begin{eqnarray}
\rho &=& \frac{1}{2}\cosh^{-1}(w_0\cosh 2\tilde{\sigma}), 
\nonumber \\
t &=& \tilde{\tau} + \tan^{-1}\left( \coth 2\rho_0 e^{2\tilde{\sigma}} 
+ \frac{1}{\sinh 2\rho_0} \right),
\nonumber \\
\phi &=& \tilde{\tau} + \tan^{-1}\left( \coth 2\rho_0 e^{2\tilde{\sigma}} 
- \frac{1}{\sinh 2\rho_0} \right)
\label{sp}\end{eqnarray}
with $\tilde{\sigma} =(L/2\pi)\sigma$,  $\tilde{\tau} =(L/2\pi)\tau$,
where $L/2\pi$ takes large value in the long string limit.
We put $\mu \equiv L/2\pi$ for convenience. The radial coordinate
$\rho$ is restricted by $\rho \ge \rho_0$ with $w_0 = \cosh 2\rho_0$.
The string shape is represented by a single arc whose endpoints approach
the boundary of $AdS_3$. The angular separation $\Delta \theta = 
2\Delta \phi$ at constant $t$ of the arcs is given by 
$\Delta \theta = 2\tan^{-1}(1/\sinh 2\rho_0)$.

Now we make the analytic continuation (\ref{an}) of the worldsheet
time for the embedding coordinates of the solution (\ref{sp})
to have
\begin{eqnarray}
Y_{-1} &=& \cosh\rho \cosh(\mu\te + i\phi_+), \;\; 
Y_0 = -i\cosh\rho \sinh(\mu\te + i\phi_+), 
\nonumber \\
Y_1 &=& \sinh\rho \cosh(\mu\te + i\phi_-), \;\; 
Y_2 = -i\sinh\rho \sinh(\mu\te + i\phi_-),
\end{eqnarray}
where $\phi_{\pm} = \tan^{-1}(\coth 2\rho_0e^{2\tilde{\sigma}}
\pm 1/\sinh 2\rho_0)$. The SO(2,4) interchange 
 produces the string configuration
in the Minkowskian Poincare coordinates. Using the following expressions
\begin{eqnarray}
\sin\phi_+ &=& \frac{\sqrt{w_0}e^{\mu\sigma} + \frac{1}{\sqrt{w_0}}
e^{-\mu\sigma} }{2\cosh\rho}, \hspace{1cm}  
\cos\phi_+ = \frac{\sqrt{w_0 - \frac{1}{w_0} }e^{-\mu\sigma} }
{2\cosh\rho}, \nonumber \\
\sin\phi_- &=& \frac{\sqrt{w_0}e^{\mu\sigma} - \frac{1}{\sqrt{w_0}}
e^{-\mu\sigma} }{2\sinh\rho}, \hspace{1cm}  
\cos\phi_- = \frac{\sqrt{w_0 - \frac{1}{w_0} }e^{-\mu\sigma} }
{2\sinh\rho}
\label{sc}\end{eqnarray}
with
\begin{equation}
\cosh\rho = \sqrt{ w_0\cosh^2\mu\sigma - \sinh^2\rho_0}, \hspace{1cm} 
\sinh\rho = \sqrt{ w_0\cosh^2\mu\sigma - \cosh^2\rho_0},
\label{cs}\end{equation}
we have
\begin{eqnarray}
\frac{1}{z} &=& \frac{1}{2}e^{-\mu \te}\left[
\sqrt{w_0 - \frac{1}{w_0} } 
e^{- \mu\sigma} - i \left( \sqrt{w_0}e^{\mu\sigma} + 
\frac{1}{\sqrt{w_0}}e^{-\mu\sigma} \right) \right],
\nonumber \\
\frac{x_0 \pm x_1}{z} &=& \pm \frac{1}{2}e^{\pm\mu \te}\left[ 
\sqrt{w_0 - \frac{1}{w_0} } e^{- \mu\sigma} \pm i \left( 
\sqrt{w_0}e^{\mu\sigma} - \frac{1}{\sqrt{w_0}}e^{-\mu\sigma} \right)
 \right].
\end{eqnarray}

The group element $g$ reads 
\begin{equation}
g = \left( \begin{array} {cc} \frac{e^{\mu\te}}{2}\left[ 
W e^{-\mu\sigma} + i \left( \sqrt{w_0}e^{\mu\sigma} - 
\frac{e^{-\mu\sigma}}{\sqrt{w_0}} \right) \right] & 
 \frac{e^{-\mu\te}}{2}\left[ W e^{-\mu\sigma} - i \left( \sqrt{w_0}
e^{\mu\sigma} + \frac{e^{-\mu\sigma}}{\sqrt{w_0}} \right) \right] \\
-\frac{e^{\mu\te}}{2}\left[ W e^{-\mu\sigma} + i \left( \sqrt{w_0}
e^{\mu\sigma} + \frac{e^{-\mu\sigma}}{\sqrt{w_0}} \right) \right] &
-\frac{e^{-\mu\te}}{2}\left[ W e^{-\mu\sigma} - i \left( 
\sqrt{w_0}e^{\mu\sigma} -
 \frac{e^{-\mu\sigma}}{\sqrt{w_0}} \right) \right]
\end{array} \right)
\label{gs}\end{equation}
with $W = \sqrt{w_0 - 1/w_0}$,
where the Euclidean worldsheet coordinates $\te$ and $\sigma$ are also
given by (\ref{st}). The determinant of $g$ is indeed unity and (\ref{gs})
is regarded as  a complexified version of the solution.

Although $g$ is expressed in an involved form, the two currents 
$j$ and $\bar{j}$ turn out to be 
\begin{eqnarray}
j &=& \left( \begin{array} {cc} -i\frac{\mu}{2} & 
-\frac{\mu}{2}e^{i\mu(w - \bar{w})} \\ -\frac{\mu}{2}
e^{-i\mu(w - \bar{w})} &  i\frac{\mu}{2} \end{array}
\right), \nonumber \\
\bar{j} &=& \left( \begin{array} {cc} i\frac{\mu}{2} & 
-\frac{\mu}{2}e^{i\mu(w - \bar{w})} \\ -\frac{\mu}{2}
e^{-i\mu(w - \bar{w})} &  -i\frac{\mu}{2} \end{array} \right), 
\end{eqnarray}
which do not include the parameter $w_0$ and are compared with
(\ref{jh}), (\ref{bj}) to yield the genus-0 algebraic curve
\begin{equation}
 y^2 = 1 - x^2.
\end{equation}
Thus the algebraic curve for the large spin limit of the spiky string
in $AdS_3$ is the same as the null cusp Wilson loop.
This agreement is associated with the SO(2,4) rotation between the 
long spiky string solution with two spikes and the straight folded string
solution, namely the GKP string solution  in the large 
spin limit \cite{KT,SRY}.

\section{Algebraic curve for the long folded string with two
spins $S, J$  and  winding number $m$}

We analyze the case for the extension of the homogeneous string 
solution in (\ref{me}), (\ref{ho}) to have one additional parameter,
a winding number $m$ of the string around the $S^1$  in $S^5$.
In the conformal gauge the asymptotic string solution is given 
by \cite{GRR}
\begin{eqnarray}
t &=& \kappa \tau, \;\; \rho = \rho(\sigma), \;\; \phi = \kappa \tau
+ \Phi(\sigma), \;\; \varphi = \nu\tau + m\sigma, 
\nonumber \\
\cosh\rho(\sigma) &=& \sqrt{1 + \gamma^2}\cosh\mu\sigma, \;\;
\tan\Phi(\sigma) = \gamma \coth\mu\sigma,   \;\;
\gamma \equiv \frac{\nu m}{\kappa \mu}
\end{eqnarray}
for the special large spin scaling limit 
\begin{eqnarray}
\kappa &\gg& 1, \;\; \mu \approx \frac{1}{\pi} \ln\frac{S}{\sqrt{\la}}
\gg 1, \;\; \nu = \frac{J}{\sqrt{\la}} \gg 1, \;\; m \gg 1,
\nonumber \\
\hat{\nu} &\equiv& \frac{\nu}{\mu} = \mathrm{fixed},
\hspace{1cm} \hat{m} \equiv \frac{m}{\mu} = \mathrm{fixed},
\end{eqnarray}
where the Virasoro constraint yields 
\begin{equation}
\kappa^2 = \mu^2 + \nu^2 + m^2.
\label{vm}\end{equation}

Making the analytic continuation  (\ref{an}) of the worldsheet time for 
the embedding coordinates in $AdS_3$ we have
\begin{eqnarray}
Y_{-1} &=& \cosh\rho \cosh\kappa\te, \hspace{1cm} 
Y_0 = -i\cosh\rho \sinh\kappa\te, \nonumber \\
Y_1 &=& \sinh\rho \cosh(\kappa\te + i\Phi), \hspace{1cm} 
Y_2 = -i\sinh\rho \sinh(\kappa\te + i\Phi),
\end{eqnarray}
which also follow the SO(2,4) interchange between $Y_0$ and $Y_2$.
We use the following expressions that are compared with (\ref{sc})
and (\ref{cs})
\begin{eqnarray}
\sin \Phi &=& \frac{\gamma\coth\mu\sigma}{\sqrt{ 1 + 
\gamma^2\coth^2\mu\sigma} }, \hspace{1cm}
\cos \Phi = \frac{1}{\sqrt{ 1 + \gamma^2\coth^2\mu\sigma} },
\nonumber \\
\sinh\rho &=& \sinh\mu\sigma\sqrt{ 1 + \gamma^2\coth^2\mu\sigma}
\end{eqnarray}
to express the string configuration in terms of the Minkowskian Poincare
coordinates 
\begin{eqnarray}
\frac{1}{z} &=& \sqrt{1 + \gamma^2}\cosh\mu\sigma e^{-\kappa\te},
\nonumber \\
\frac{x_0}{z} &=& \sinh\mu\sigma \sinh\kappa\te +i\gamma 
\cosh\mu\sigma \cosh\kappa\te, \nonumber \\  
\frac{x_1}{z} &=& \sinh\mu\sigma \cosh\kappa\te +i\gamma 
\cosh\mu\sigma \sinh\kappa\te.
\end{eqnarray}
The corresponding group element $g$ is described by
\begin{equation}
g = \left( \begin{array}{cc} e^{\kappa\te}( \sinh\mu\sigma
+ i\gamma \cosh\mu\sigma) & \sqrt{1 + \gamma^2}e^{-\kappa\te}
\cosh\mu\sigma \\ -\sqrt{1 + \gamma^2}e^{\kappa\te}
\cosh\mu\sigma & e^{-\kappa\te}( -\sinh\mu\sigma
+ i\gamma \cosh\mu\sigma) \end{array} \right),
\label{mg}\end{equation}
which is also regarded as a complexified version of the 
solution. The currents $j$ and $\bar{j}$ are computed by
\begin{eqnarray}
j &=& \left( \begin{array}{cc} -\frac{i}{2}(\kappa - \mu\gamma) &
\frac{\mu}{2}\sqrt{1 + \gamma^2}e^{i\kappa(w-\bar{w})} \\ 
\frac{\mu}{2}\sqrt{1 + \gamma^2}e^{-i\kappa(w-\bar{w})} & 
\frac{i}{2}(\kappa - \mu\gamma) \end{array} \right), 
\nonumber \\
\bar{j} &=& \left( \begin{array}{cc} \frac{i}{2}(\kappa + \mu\gamma) &
\frac{\mu}{2}\sqrt{1 + \gamma^2}e^{i\kappa(w-\bar{w})} \\ 
\frac{\mu}{2}\sqrt{1 + \gamma^2}e^{-i\kappa(w-\bar{w})} & 
-\frac{i}{2}(\kappa + \mu\gamma) \end{array} \right).
\label{jm}\end{eqnarray}

The two independent solutions for the linear problem are determined by
\begin{eqnarray}
\Psi_1 &=& e^{-\frac{\kappa}{2}\sqrt{d}\left( \frac{1}{1-x}w +  
\frac{1}{1+x}\bar{w} \right)} \left( \begin{array}{c} 
e^{i\frac{\kappa}{2}(w - \bar{w})} \\
\frac{\kappa}{\mu\sqrt{1 + \gamma^2}}\left(\sqrt{d} +
 i(x - \frac{\mu\gamma}{\kappa})\right) 
e^{-i\frac{\kappa}{2}(w - \bar{w})} \end{array}\right), 
\label{pf} \\
\Psi_2 &=& e^{\frac{\kappa}{2}\sqrt{d}\left( \frac{1}{1-x}w +  
\frac{1}{1+x}\bar{w} \right)}\left( \begin{array}{c} 
e^{i\frac{\kappa}{2}(w - \bar{w})} \\
\frac{\kappa}{\mu\sqrt{1 + \gamma^2}}\left(-\sqrt{d} +
 i(x - \frac{\mu\gamma}{\kappa}) \right) 
e^{-i\frac{\kappa}{2}(w - \bar{w})} \end{array}\right),  
\label{ps}\end{eqnarray}
where 
\begin{equation}
d = \left(\frac{\mu}{\kappa}\right)^2 + \frac{2\mu\gamma}{\kappa}x
- x^2.
\end{equation}
When $\gamma = 0$ these solutions become (\ref{so}) and (\ref{ts}).
Arranging the two independent solutions to construct $\hat{\Psi}$
and choosing $A = \sqrt{d}\mathrm{diag}(1,-1)$ we obtain the
polynomial Lax matrix
\begin{equation}
L(w,\bar{w};x) = \left( \begin{array} {cc} -i\left(x - 
\frac{\mu\gamma}{\kappa}\right) & \frac{\mu}{\kappa} \sqrt{1 + \gamma^2}
e^{i\kappa(w-\bar{w})} \\ \frac{\mu}{\kappa}\sqrt{1 + \gamma^2}
e^{-i\kappa(w-\bar{w})} & i\left(x - \frac{\mu\gamma}{\kappa}\right)
\end{array} \right),
\label{lm}\end{equation}
which gives the genus-0 algebraic curve $\Sigma$
\begin{equation}
y^2 = \left(\frac{\mu}{\kappa}\right)^2 + \frac{2\mu\gamma}{\kappa}x 
- x^2.
\label{cm}\end{equation}
The two branch points of the curve on the $x$ plane read
\begin{equation}
-1 < x^-= -\frac{\mu}{\kappa} (\sqrt{1 + \gamma^2} - \gamma ) < 0,
\hspace{1cm}
0 < x^+= \frac{\mu}{\kappa} (\sqrt{1 + \gamma^2} + \gamma ) < 1,
\label{br}\end{equation}
where inequality is shown by using the Virasoro constraint (\ref{vm}).
In terms of the fixed scaling parameters $\hat{\nu}$ and $\hat{m}$
the algebraic curve is expressed as
\begin{equation}
y^2 = \frac{1}{1 + \hat{\nu}^2 + \hat{m}^2} + \frac{2\hat{\nu}\hat{m}}
{1 + \hat{\nu}^2 + \hat{m}^2}x - x^2,
\end{equation}
whose branch points are given by
\begin{equation}
x^{\pm} = \frac{\hat{\nu}\hat{m} \pm 
\sqrt{(1 + \hat{\nu}^2)(1 + \hat{m}^2)} }{ 1 + \hat{\nu}^2 + \hat{m}^2}.
\end{equation}
When the winding number $m$ is turned off, the algebraic curve
reduces to (\ref{ch}).

Now following the method in ref. \cite{JG} we reconstruct the solutions
of linear problem from the algebraic curve $\Sigma$ (\ref{cm}) and some
assumptions on the form of $L$. The reconstruction is performed by using
the fact that since the linear operators $\pa + J$ and
$\bar{\pa} + \bar{J}$ can be simultaneously diagonalized with 
the monodromy operator $\Omega$, 
a solution of the linear problem should be proportional
to the normalized eigenvector $\Psi_n$ of $L$
\begin{eqnarray}
\Psi(w,\bar{w};x) &=& f_{BA}(w,\bar{w};x)\Psi_n(w,\bar{w};x),
\nonumber \\
\Psi_n(w,\bar{w};x) &=& \left(\begin{array}{c} 1 \\ \psi_n(w,\bar{w};x)
\end{array} \right),
\end{eqnarray}
where $f_{BA}(w,\bar{w};x)$ is a scalar function called as the 
Baker-Akhiezer function.  The flat currents $J$ and $\bar{J}$
can be extracted by the singular terms in the Laurent expansion of some
polynomial in $L$ with coefficients being rational functions of $x$.
For the Lax operator that is polynomial in $x$  this extraction
procedure is expressed as
\begin{equation}
\left[\frac{c_1}{1 - x} L(w,\bar{w};x)\right]_{x=1}^- = J(w,\bar{w};x),
\;\; \left[\frac{c_{-1}}{1 + x} L(w,\bar{w};x)\right]_{x=-1}^- = 
\bar{J}(w,\bar{w};x),
\label{cl}\end{equation}
so that the solution of the linear problem has the following asymptotic
behavior near the essential singularities $x = \pm1$
\begin{equation}
\Psi \approx e^{-\frac{c_1 y(x)}{1-x}w -\frac{c_{-1} y(x)}{1+x}\bar{w} }.
\label{as}\end{equation}

The substitution of (\ref{jm}) and (\ref{lm}) into (\ref{cl}) provides
\begin{equation}
c_1 = c_{-1} = \frac{\kappa}{2},
\end{equation}
which combines with the starting spectral curve $\Sigma$ (\ref{cm}) and
(\ref{as}) to yield
\begin{equation}
\Psi \approx e^{-\frac{\kappa}{2}\sqrt{  \left(\frac{\mu}{\kappa}\right)^2
 + \frac{2\mu\gamma}{\kappa}x - x^2 }\left( \frac{1}{1-x}w + 
\frac{1}{1+x}\bar{w} \right)  },
\label{ex}\end{equation}
which is a part of the Baker-Akhiezer function and agrees with the overall
exponential factor in (\ref{pf}). As shown in (\ref{br}) the algebraic 
curve has two branch points $x^{\pm}$ inside the interval $[-1,1]$ 
in the $x$ plane.  When $\gamma=0$ and $\mu=\kappa$,
the branch points $x^+$ and $x^-$ collide with the essential singular
points 1 and $-1$ respectively. 

We uniformize the algebraic curve $\Sigma$ (\ref{cm}) as
\begin{equation}
y = \frac{\mu \sqrt{1 + \gamma^2}}{\kappa} \frac{2t}{1 + t^2},
\hspace{1cm} x = \frac{\mu \sqrt{1 + \gamma^2}}{\kappa}
 \frac{1 - t^2}{1 + t^2} + \frac{\mu \gamma}{\kappa},
\end{equation}
where passing from  one sheet to the other one is performed by the
transformation $t \rightarrow -t$. The two points $x = \infty^+$ and 
$x = \infty^-$ in the algebraic curve  $\Sigma$ (\ref{cm}) above
$x = \infty$ in the complex $x$ plane are provided by $t = -i$ and
$t = i$ respectively. The two points above $x = 1$ correspond to
\begin{equation}
t = \pm \sqrt{ \frac{\mu \sqrt{1 + \gamma^2} - (\kappa - \mu \gamma)}
{\mu \sqrt{1 + \gamma^2} + \kappa - \mu \gamma}  } = \pm t_1,
\end{equation}
where the imaginary  $t_1 = i\sqrt{(1-x^+)/(1-x^-)}$ reduces
to zero for $\gamma = 0, \; \kappa=\mu$, and $|t_1| < 1$,
while those above $x = -1$ correspond to
\begin{equation}
t = \pm \sqrt{ \frac{\mu \sqrt{1 + \gamma^2} + \kappa + \mu \gamma}
{\mu \sqrt{1 + \gamma^2} - (\kappa + \mu \gamma)}  } = \pm t_{-1},
\end{equation}
where the imaginary  $t_{-1} = -i\sqrt{(1+x^+)/(1+ x^-)}$ becomes
infinte for $\gamma = 0, \; \kappa=\mu$, and $1 < |t_{-1}|$.

The Lax matrix is diagonal at $x = \infty$, which leads to
\begin{equation}
\Psi_n(w,\bar{w};x = \infty^+) = \left(\begin{array}{c} 1 \\ 0 
\end{array} \right), \;\;
\Psi_n(w,\bar{w};x = \infty^-) = \left(\begin{array}{c} 1 \\ \infty 
\end{array} \right),
\end{equation}
where $\Psi_n$ has a single pole at $x = \infty^-$, equivalently 
$t = i$, since the genus of the algebraic curve (\ref{cm}) is zero
such that there is no dynamical divisor. Therefore the normalized 
eigenvector  can be parameterized by
\begin{equation}
\Psi_n(w,\bar{w};x ) = \left(\begin{array}{c} 1 \\ 
a(w,\bar{w}) \frac{t + i}{t - i} \end{array} \right).
\label{nv}\end{equation}
For  one sheet, $t$ is given by
\begin{equation}
t = \sqrt{ \frac{\mu \sqrt{1 + \gamma^2} - (\kappa x- \mu \gamma)}
{\mu \sqrt{1 + \gamma^2} + \kappa x - \mu \gamma}  } 
= \sqrt{\frac{x^+ - x}{x - x^-}},
\end{equation}
whose substitution into the lower component of (\ref{nv}) yields
\begin{equation}
ia(w,\bar{w})\frac{\kappa}{\mu \sqrt{1 + \gamma^2}} \left( \sqrt{d} +
i\left( x - \frac{\mu \gamma}{\kappa} \right) \right),
\end{equation}
that is compared with (\ref{pf}), while for the other sheet the 
sign-changed expression $t = -\sqrt{(x^+ - x)(x - x^-)}$ leads to
\begin{equation}
-ia(w,\bar{w})\frac{\kappa}{\mu \sqrt{1 + \gamma^2}} \left( \sqrt{d} -
i\left( x - \frac{\mu \gamma}{\kappa} \right) \right),
\end{equation}
that is compared with (\ref{ps}).
The exponent of (\ref{ex}) is expressed  in terms of $t$ as
\begin{equation}
\exp \left[-\frac{\kappa\mu \sqrt{1 + \gamma^2}}{\mu \sqrt{1 + \gamma^2}
+ \kappa - \mu \gamma} \frac{t}{t^2-t_1^2}w + 
\frac{\kappa\mu \sqrt{1 + \gamma^2}}{\mu \sqrt{1 + \gamma^2}
- ( \kappa + \mu \gamma)} \frac{t}{t^2-t_{-1}^2}\bar{w} \right]
\equiv c(w,\bar{w};t).
\label{ct}\end{equation}
Thus we have 
\begin{eqnarray} 
\Psi(w,\bar{w};t) &=& f_{BA}(w,\bar{w};t)\Psi_n(w,\bar{w};t)
\nonumber \\
&=& c(w,\bar{w};t)b(w,\bar{w})\left(\begin{array}{c} 1 \\ 
a(w,\bar{w}) \frac{t + i}{t - i} \end{array} \right).
\end{eqnarray}

Since the flat currents vanish in $x = \infty$, the solution 
$\Psi(w,\bar{w};t)$ of the linear problem becoms $w$ and $\bar{w}$
independent. At $x = \infty^+, \; (t =-i)$ the upper component
$c(w,\bar{w};-i)b(w,\bar{w})$ should be constant so that 
\begin{equation}
b(w,\bar{w}) = \frac{1}{c(w,\bar{w};-i)} = e^{i\frac{\kappa}{2}
(w - \bar{w})},
\end{equation}
which is evaluated from (\ref{ct}). 
 At $x = \infty^-, \; (t = i)$ the $w$ and $\bar{w}$ independence
of the lower component also yields
\begin{equation}
b(w,\bar{w})a(w,\bar{w}) = \frac{1}{c(w,\bar{w};i)} = 
e^{-i\frac{\kappa}{2}(w - \bar{w})}.
\end{equation}
From the algebraic curve (\ref{cm}) we have reconstructed two 
independent solutions of the linear problem, (\ref{pf}) and (\ref{ps}).
Owing to (\ref{gx}) these solutions give the following group element which
we denote by $\hat{g}$
\begin{equation}
\hat{g} = -i \frac{(1 + \gamma^2)^{1/4}}{\sqrt{2} } \left( 
\begin{array}{cc} -\frac{1 + i\gamma}{\sqrt{1 + \gamma^2}}
e^{\kappa \te + \mu\sigma} & -e^{-\kappa \te + \mu\sigma} \\
-\frac{1 - i\gamma}{\sqrt{1 + \gamma^2}}e^{\kappa \te - \mu\sigma} &
e^{-\kappa \te - \mu\sigma} \end{array} \right).
\end{equation}
This expression seems to be different from the original group
element $g$ (\ref{mg}), but there is 
a freedom under the left multiplication
by the $w$ and $\bar{w}$ independent group element. Indeed we
produce the original string configuratin by the following 
constant rotation
\begin{equation}
g = -\frac{i}{\sqrt{2}} \left( \begin{array}{cc} (1 + \gamma^2)^{1/4} &
-(1 + \gamma^2)^{1/4} \\ -\frac{1 - i\gamma}{(1 + \gamma^2)^{1/4} }   &
 -\frac{1 + i\gamma}{(1 + \gamma^2)^{1/4} } \end{array} \right)
\hat{g}.
\end{equation} 

\section{Algebraic curve for the circular winding string in 
$AdS_3\times S^1$}

We consider the circular spinning string solution with spins and
winding numbers $(S, n)$ in $AdS_3$ and $(J,m)$ in $S^1$ \cite{ART}.
The string configuration in $AdS_3$ is described by the embedding 
coordinates
\begin{equation}
Y_{-1} + iY_0 = \cosh\rz e^{it}, \hspace{1cm} Y_1 + iY_2 = 
\sinh\rz e^{i\phi}
\end{equation}
with $t = \kappa\tau, \phi = \omega\tau + n\sigma$, and the $S^1$ part is
given by $\varphi = \nu \tau - m\sigma$ with $J = \nu \sqrt{\la}$
in the metric (\ref{me}).
The Virasoro constraint yields
\begin{equation}
-\kappa^2\cosh^2\rz + ( \omega^2 + n^2 )\sinh^2\rz + \nu^2 + m^2 = 0,
\;\; nS = mJ.
\end{equation}

Here we also rotate the worldsheet time as (\ref{an}) and make the 
interchange between $Y_0$ and $Y_2$ to compute the group element
\begin{equation}
g = \left( \begin{array}{cc} \sinh\rz e^{\omega\te + in\sigma} &
\cosh\rz e^{-\kappa\te} \\ -\cosh\rz e^{\kappa\te} &
-\sinh\rz e^{-(\omega\te + in\sigma)} \end{array} \right),
\end{equation}
which is expressed through (\ref{st}) as
\begin{equation}
g = \left( \begin{array}{cc} \sinh\rz e^{-\frac{i}{2}(\ob w 
- \oa\bar{w})}& \cosh\rz e^{\frac{i}{2}\kappa(w - \bar{w}) } \\
-\cosh\rz e^{-\frac{i}{2}\kappa(w - \bar{w}) } & 
-\sinh\rz e^{\frac{i}{2}(\ob w - \oa\bar{w})} \end{array} \right) 
\end{equation}
with $\omega_{\pm} = \omega \pm n$. The currents $j$ and $\bar{j}$ are
obtained by
\begin{eqnarray}
j &=& \frac{i}{2}\left( \begin{array}{cc} \ob \sinh^2\rz - 
\kappa\cosh^2\rz & (\ob - \kappa)H e^{\frac{i}{2}
((\ob  + \kappa)w - (\oa  + \kappa)\bar{w}) } \\
-(\ob - \kappa)H e^{-\frac{i}{2}
((\ob  + \kappa)w - (\oa  + \kappa)\bar{w}) } &
-(\ob \sinh^2\rz - \kappa\cosh^2\rz ) \end{array} \right),
\nonumber \\
\bar{j} &=& -\frac{i}{2}\left( \begin{array}{cc} \oa \sinh^2\rz - 
\kappa\cosh^2\rz & (\oa - \kappa)H e^{\frac{i}{2}
((\ob  + \kappa)w - (\oa  + \kappa)\bar{w}) } \\
-(\oa - \kappa)H e^{-\frac{i}{2}
((\ob  + \kappa)w - (\oa  + \kappa)\bar{w}) } &
-(\oa \sinh^2\rz - \kappa\cosh^2\rz ) \end{array} \right)
\end{eqnarray}
with $H = \sinh\rz \cosh\rz $.

The first solution $\Psi_1$ for the linear problem is  expressed as
\begin{equation}
\Psi_1 = e^{-\frac{\sqrt{d}}{4} \left(
\frac{\ob + \kappa}{1-x}w + \frac{\oa + \kappa}{1+x}\bar{w} \right) }
\left(\begin{array}{c} e^{\frac{i}{4}((\ob +\kappa)w - 
(\oa +\kappa)\bar{w}) } \\  e^{-\frac{i}{4}((\ob +\kappa)w - 
(\oa +\kappa)\bar{w}) }f \end{array} \right).
\end{equation}
From the differential equations in (\ref{al}) and (\ref{be}) $f$ and $d$ 
are determined by
\begin{eqnarray}
d &=& - \left( \frac{\ob - \kappa}{\ob + \kappa} \right)^2 +
2\cosh2\rz \frac{\ob - \kappa}{\ob + \kappa}x - x^2 \equiv d_-,
\nonumber \\
f &=& -\frac{i}{\sinh 2\rz} \frac{\ob + \kappa}{\ob - \kappa} 
\left( \sqrt{d_-} +i \left( x - \cosh 2\rz 
\frac{\ob - \kappa}{\ob + \kappa} \right) \right),
\end{eqnarray}
while the differential equations in (\ref{bb}) lead to
\begin{eqnarray}
d &=& - \left( \frac{\oa - \kappa}{\oa + \kappa} \right)^2 -
2\cosh2\rz \frac{\oa - \kappa}{\oa + \kappa}x - x^2 \equiv d_+,
\nonumber \\
f &=& \frac{i}{\sinh 2\rz} \frac{\oa + \kappa}{\oa - \kappa} 
\left( \sqrt{d_+} +i \left( x + \cosh 2\rz 
\frac{\oa - \kappa}{\oa + \kappa} \right) \right).
\end{eqnarray}
However, there is a relation between $\omega, \kappa$ and
$n$ \cite{ART}
\begin{equation}
\kappa^2 = \omega^2 - n^2 = \oa \ob,
\end{equation}
so that $d_+ = d_-$ and the two expressions are identical.
The second solution $\Psi_2$ for the linear problem
is also constructed by
\begin{equation}
\Psi_2 = e^{\frac{\sqrt{d}}{4} \left(
\frac{\ob + \kappa}{1-x}w + \frac{\oa + \kappa}{1+x}\bar{w} \right) }
\left(\begin{array}{c} e^{\frac{i}{4}((\ob +\kappa)w - 
(\oa +\kappa)\bar{w}) } \\  e^{-\frac{i}{4}((\ob +\kappa)w - 
(\oa +\kappa)\bar{w}) }h \end{array} \right),
\end{equation}
where $h$ has the corresponding two equivalent expressions
that we call as the  minus and plus expressions
\begin{eqnarray}
h &=& \frac{i}{\sinh 2\rz} \frac{\ob + \kappa}{\ob - \kappa} 
\left( \sqrt{d_-} -i \left( x - \cosh 2\rz 
\frac{\ob - \kappa}{\ob + \kappa} \right) \right),
\nonumber \\
h &=& -\frac{i}{\sinh 2\rz} \frac{\oa + \kappa}{\oa - \kappa} 
\left( \sqrt{d_+} -i \left( x + \cosh 2\rz 
\frac{\oa - \kappa}{\oa + \kappa} \right) \right).
\end{eqnarray}

We use the minus expression to construct the polynomial Lax matrix
by choosing $A =\sqrt{d_-}\mathrm{diag}(1,-1)$
\begin{equation}
L = \left( \begin{array}{cc} -i\left( x - \cosh 2\rz 
\frac{\ob - \kappa}{\ob + \kappa} \right) & i\sinh2\rz 
\frac{\ob - \kappa}{\ob + \kappa}e^{\frac{i}{2}((\ob +\kappa)w - 
(\oa +\kappa)\bar{w}) } \\
-i\sinh2\rz \frac{\ob - \kappa}{\ob + \kappa}e^{-\frac{i}{2}
((\ob +\kappa)w - (\oa +\kappa)\bar{w}) } &
i\left( x - \cosh 2\rz \frac{\ob - \kappa}{\ob + \kappa} \right)
\end{array} \right).
\end{equation}
The corresponding algebraic curve  is obtained by
\begin{equation}
y^2 = \sinh^2 2\rz \left( \frac{\ob - \kappa}{\ob + \kappa} \right)^2
- \left( x - \cosh 2\rz \frac{\ob - \kappa}{\ob + \kappa} \right)^2,
\end{equation}
while the equivalent curve in the plus expression is given by
\begin{equation}
y^2 = \sinh^2 2\rz \left( \frac{\oa - \kappa}{\oa + \kappa} \right)^2
- \left( x + \cosh 2\rz \frac{\oa - \kappa}{\oa + \kappa} \right)^2.
\end{equation}
Thus we have a single expression of the genus-0 algebraic curve 
\begin{equation}
y^2 = - \left( \frac{n}{\omega + \kappa} \right)^2 - 2\cosh2\rz \left( 
\frac{n}{\omega + \kappa} \right)x - x^2.
\end{equation}
For $n > 0$ there are two branch points such that $-1 < x^- < x^+ < 0$,
\begin{equation}
x^+ = -\frac{n}{\omega + \kappa}e^{-2\rz}, \hspace{1cm} 
x^- = -\frac{n}{\omega + \kappa}e^{2\rz},
\end{equation}
where $\kappa\sqrt{\la}$ has the $\la/J^2$ expansion \cite{ART}
\begin{equation}
\kappa \sqrt{\la} = J +  \frac{\la}{2J^2}\left( m^2J + 2n^2 S \right)
+ \cdots
\end{equation}
with $S/J = m/n$, and $n/( \omega + \kappa)$ is expanded as 
\begin{equation}
\frac{n}{\omega + \kappa} = \frac{n\sqrt{\la}}{2J} \left( 1 - 
\frac{\la}{4J^2}\left( n^2 + 2m^2 + 4n^2\frac{S}{J} \right) 
+ \cdots \right).
\end{equation}
We use the $\la/J^2$ expansion for $\cosh^2\rz$ \cite{SR}
\begin{equation}
\cosh^2\rz = 1 + \frac{S}{J} - \frac{\la}{2J^2}\left( (n^2 + m^2 )
\frac{S}{J} + 2n^2\frac{S^2}{J^2} \right) + \cdots
\end{equation}
to express $e^{\pm2\rz}$ as
\begin{eqnarray}
e^{\pm 2\rz} &=& 2\left( \cosh^2\rz - \frac{1}{2} \pm \cosh\rz 
\sqrt{\cosh^2\rz - 1} \right)  \\
&=& 1 + \frac{2S}{J} \pm 2\sqrt{\frac{S}{J} + \frac{S^2}{J^2} }
- \frac{\la}{2J^2} \left( 2 \pm \frac{ 1 + \frac{2S}{J}}
{\sqrt{\frac{S}{J}  + \frac{S^2}{J^2} } } \right)
 \left( (n^2 + m^2 )\frac{S}{J} + 2n^2\frac{S^2}{J^2} 
\right) + \cdots. \nonumber
\end{eqnarray}

\section{Conclusion}

Based on the algebraic curve prescription \cite{JG} we have solved the 
linear problems for the analytic continued and SO(2,4) rotated string
cnfigurations of the homogeneous string solutions that are given by 
the special large spin scaling limits of the folded string solution
with spins $S$ and $J$ in $AdS_3\times S^1$ \cite{FTT} as well as of 
the generalized string solution with an additional winding number $m$ in
$S^1$ \cite{GRR}. Combining two independent solutions of the linear
problems to construct the polynomial Lax operators, we have derived the
algebraic curves as the second-order polynomials of the spectral
variable $x$ with coefficients expressed by the relevant fixed
scaling parameters.

We have observed that the locations of two branch points 
of the algebraic curve are
so controlled by the Virasoro constraint as to stay between
the essential singular points $x = \pm1$ in the $x$ plane. 
It has been shown that
in the $J = m = 0$ case, namely the large spin GKP string case
the branch points collide with the essential singularities so that
the algebraic curve reduces to that for the null cusp Wilson loop.

Inversely, starting from the obtained algebraic curve we have
reconstructed the generalized homogeneous string configuration 
by using some assumptions about the Lax operator and making
an appropriate constant rotation.

The algebraic curve for the large spin limit
of the spiky string solution in $AdS_3$ \cite{DL} has been
computed and it has been shown 
that the large spin spiky string and the large spin
GKP string have the same algebraic  curve as the null cusp Wilson
loop. These coincidences are regarded as the reflections of the
link \cite{KRT} between the large spin GKP string and the null 
cusp Wilson loop as well as the link \cite{KT,SRY}
between the large spin GKP string and the
large spin spiky string through the SO(2,4) transformations 
accompanied with the analytic continuations.

We have demonstrated that the algebraic curve for the circular
winding string solution with spins $(S, J)$ and winding numbers
$(n,m)$ in $AdS_3\times S^1$ \cite{ART} 
becomes the genus-0 curve, that is,
the second-order polynomial including a linear term similar to
that for the generalized homogeneous string configuration with
spins  $(S, J)$  and winding number $m$ in $AdS_3\times S^1$.
We have observed that the latter has a reduction such that 
the linear term vanishes for $m = 0$, while the former is not
allowed to take such a reduction and has two branch points
inside the interval $[-1,1]$  owing to the
Virasoro constraint.

\end{document}